\documentstyle[aps,epsf,psfig,twocolumn]{revtex}
\begin{document}
\draft
\flushbottom
\twocolumn[
\hsize\textwidth\columnwidth\hsize\csname @twocolumnfalse\endcsname

%\draft
\title{
Superconducting and pseudogap phases from scaling near 
a Van Hove singularity}
\author{J. Gonz\'alez \\}
\address{
        Instituto de Estructura de la Materia. 
        Consejo Superior de Investigaciones Cient{\'\i}ficas. 
        Serrano 123, 28006 Madrid. Spain.}
\date{\today}
\maketitle
\widetext
\begin{abstract}
We study the quantum corrections to the Fermi energy of a 
two-dimensional electron system, showing that it is attracted 
towards the Van Hove singularity for a certain range of doping 
levels. The scaling of the Fermi level allows to cure the infrared
singularities left in the BCS channel after renormalization of
the leading logarithm near the divergent density of
states. A phase of $d$-wave superconductivity arises beyond the
point of optimal doping corresponding to the peak of the
superconducting instability. For lower doping levels, the  
condensation of particle-hole pairs due to the nesting of the 
saddle points takes over, leading to the opening of a gap
for quasiparticles in the neighborhood of the singular points.

\end{abstract}

]
\narrowtext 
\tightenlines%\newpage

The study of the electronic properties of the cuprates 
represents nowadays a great challenge from the theoretical 
point of view, as the phenomenology of these materials has 
become increasingly rich during the last decade. 
There have only been a few attempts to develop a theory
that may encompass the main experimental 
features\cite{dago}, including
the antiferromagnetism of the undoped compounds and the 
pseudogap phase above the superconducting transition. 
Additionally, other proposals have focused on the mechanism
of superconductivity, stressing the role played by 
antiferromagnetic fluctuations\cite{fluc} or by the 
proximity to a Van Hove singularity (VHS) in the doped 
materials\cite{newns}.

The later approach has received much attention recently, 
since it establishes a natural competition between 
magnetic and superconducting instabilities in a
two-dimensional (2D) system\cite{jpn,ren,prl,ren2,binz}. 
The investigation of the model of electrons near a VHS 
is delicate due to the appearance of logarithmic 
divergences in perturbation theory.
In a renormalization group (RG) framework, one has to
handle infrared singularities which arise after 
renormalizing away the leading logarithm, as the energy 
dependence of some quantities comes in powers of a 
logarithm square\cite{furu}.

Most part of the analyses of the problem have been made 
fixing the Fermi level at the VHS 
from the start. This questions the naturalness
of the predicted instabilities, that rely critically on the 
proximity to the singular density of states. The Fermi energy
is actually a dynamical quantity that is shifted by quantum
corrections. It has been shown that the VHS has the tendency 
to attract the Fermi level of the electron 
system\cite{mark,epl,george,charge}. 
It is therefore more appropriate to let the 
chemical potential free to evolve as the states are integrated
in the quantum theory. This also solves at once the problem
of the infrared divergences, as the shift of the chemical
potential from the VHS acts as an infrared
cutoff in the logarithmic dependences left in the 
renormalization.

We illustrate the above ideas in the case of the $t-t'$
Hubbard model, which has a dispersion relation with saddle 
points at $A = (\pi, 0)$ and $B = (0, \pi)$. 
We deal with a wilsonian RG approach in which the chemical
potential $\mu $ is originally placed away from the VHS, 
and electron modes in two thin slices about energies 
$\mu + \Lambda $ and $\mu - \Lambda $ are progressively 
integrated out\cite{shankar}. At each RG step, the chemical
potential is free to reaccomodate due to the self-energy
corrections from the charge integrated out.

The behavior of $\mu $ as $\Lambda \rightarrow 0$ can be 
obtained by solving the frequency and
momentum-independent part of the Schwinger-Dyson equation  
$1/G = 1/G_0 - \Sigma $. 
The charge in the slice $d\Lambda $ is given in terms of
the density of states $n(\varepsilon)$ by $n(\mu - \Lambda)
d \Lambda $. It couples through the forward-scattering vertex
$F$ in the usual Hartree and exchange diagrams. To improve
the perturbative approach, one has to
take into account that $F$ is also a scale-dependent 
quantity\cite{prl}, 
whose value vanishes in the proximity to the VHS
according to the expression
$F(\varepsilon ) \approx F_0/(1 - F_0 \log (\varepsilon )/
(4\pi^2 t))$. Thus, upon integration of high-energy modes
in a thin slice $d\Lambda $, the chemical potential $\mu $
is shifted according to
\begin{equation}
\frac{d\mu}{d\Lambda } =  
    F (\mu /(\mu - \Lambda )) \; 
   n(\mu - \Lambda) 
\label{diff}
\end{equation}

The differential renormalization given by Eq. (\ref{diff}) leads
to the scaling of $\mu $ as a function of $\Lambda $. It can
be seen that, in a certain range, 
the flow of the chemical potential is attracted by the VHS
as $\Lambda \rightarrow 0$. If $F$ were constant,
the fixed-point condition $-1 + F n(\tilde{\mu}) = 0$, 
with $\tilde{\mu} = \mu - \Lambda $, would give 
$\mu (\Lambda )$. Taking into account the scaling of the 
$F$ vertex, it is found that the renormalized value of $\mu $ 
lies very close to to the level of the singularity, 
for appropriate values of the bare chemical potential. 

\begin{figure}  
\epsfxsize=\hsize %\epsfysize = 4cm
\centerline{\epsfbox{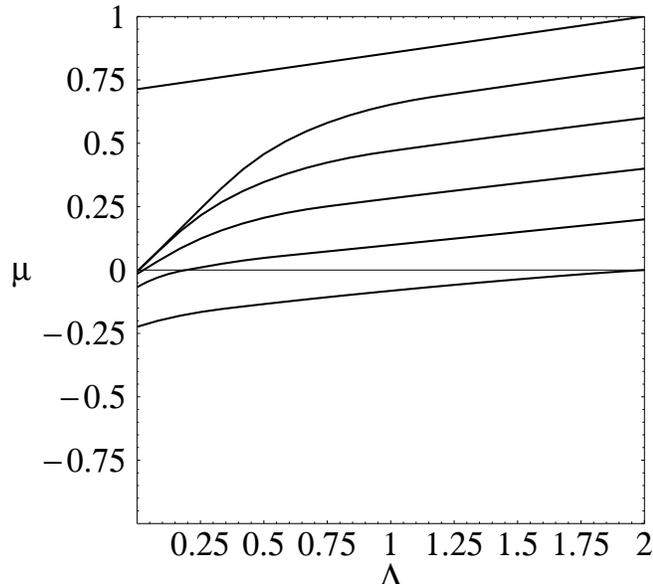}} %\vspace{-1.5cm}
\caption{Scaling of the chemical potential as a function of the
high-energy cutoff. The results correspond to the Hubbard 
coupling $U = 4t$.}
\label{one}  
\end{figure}

We have represented in Fig. \ref{one} the results of solving the
scaling equation with a model density of states 
$n(\varepsilon) = \log (t/\varepsilon) /(4\pi^2 t)$ for 
$ \left|\varepsilon \right| \leq 0.5t $, and constant elsewhere.
We observe that there is a range of nominal values of the chemical
potential in which this is attracted towards the VHS. 
As a consequence of that, there is a range of 
filling levels that are forbidden above the level of
the singularity.

In an open system, the pinning of the Fermi level to the VHS can 
be realized by exchanging particles with the environment, as 
greater stability is attained then in the system\cite{charge}. 
In a closed system,
however, the mechanism of pinning can be only realized if the 
Fermi line deforms to become closer to the saddle points of 
the dispersion relation. This is consistent with the 
renormalization of the Fermi line observed in numerical studies 
of the Hubbard model, which is translated into a shift of the 
next-to-nearest-neighbor hopping $t'$ \cite{duffy,ogata}.
Such an effect is also in agreement with recent 
experimental observations reported in Ref. \onlinecite{kord}.
It has been shown there that the Fermi line of Bi-2212 bends
progressively as the doping level is increased, so that even in
the overdoped regime it does not lose its hole-like character.

The scaling of the chemical potential $\mu (\Lambda )$
towards the VHS triggers the different instabilities in the
system. These can be traced back to the divergent behavior
of some of the interaction channels.
Among all the kinematics, two different channels 
develop a singularity for an imaginary value of the frequency,
namely the BCS channel
of colliding particles with zero total momentum and the
channel with momentum transfer ${\bf Q} \equiv (\pi, \pi)$.
Such features are related to a phenomenon of condensation and
the opening of new phases, while the singularities 
in the rest of the channels arise for real values of the
frequency and they correspond to the appeareance of new 
excited states.

\begin{figure}
\epsfxsize=7cm %\epsfysize = 4cm
\centerline{\epsfbox{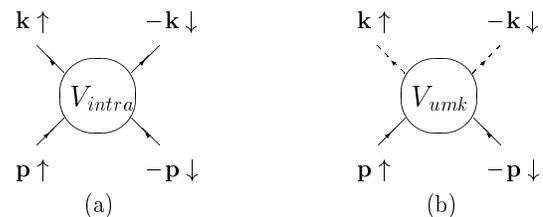}} %\vspace{-1.5cm}
\caption{BCS vertices that undergo renormalization by
particle-particle diagrams. The 
solid and dashed lines stand for modes in the neighborhood 
of the two different saddle points.}
\label{two}
\end{figure}

\begin{figure}
\epsfxsize=7cm %\epsfysize = 4cm
\centerline{\epsfbox{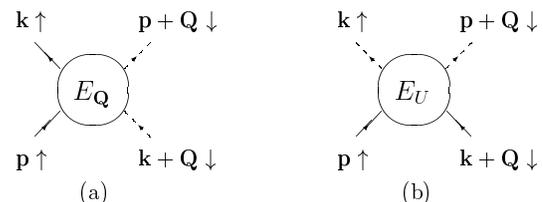}} %\vspace{-1.5cm}
\caption{Exchange and Umklapp vertices that undergo 
renormalization by particle-hole diagrams.}
\label{twop}   
\end{figure}

In what follows, we concentrate on the renormalization of the 
vertices with BCS kinematics, listed in Fig. \ref{two}, and 
of those with momentum transfer ${\bf Q} \equiv (\pi, \pi)$,
listed in Fig. \ref{twop}. In a wilsoninan RG approach of the
kind developed in Ref. \onlinecite{shankar}, processes with 
the latter kinematics are disentangled from the divergences 
in the BCS channel, at least at the one-loop level. 
This holds in particular for   
a saddle-point dispersion relation since, for kinematics in  
which the total momentum of the incoming particles is not
precisely zero, particle-particle
diagrams give irrelevant contributions $\sim (d\Lambda )^2$
at each RG step\cite{shankar}.

The converse statement is also true, and the vertices 
with the BCS kinematics $V_{intra}$ and $V_{umk}$ only 
get renormalized between themselves at the one-loop level.
This comes from the fact that the integration of modes in 
the differential slices of energy $\pm \Lambda $ only 
produces contributions of order $\sim d\Lambda$ through 
particle-particle corrections to the above mentioned 
vertices\cite{shankar}. The scaling equations for 
$V_{intra}$ and $V_{umk}$ take the form
\begin{eqnarray}
\Lambda \frac{\partial V_{intra}}{\partial \Lambda}
   & = &    c \; n(\mu - \Lambda ) \;
     \left( V_{intra}^2 + V_{umk}^2 \right)   \label{iu1}  \\
\Lambda \frac{\partial V_{umk}}{\partial \Lambda}
   & = &   2  c \; n(\mu - \Lambda ) \;
     V_{intra} V_{umk}
\label{iu2}
\end{eqnarray}
with $c \equiv 1/\sqrt{1 - 4(t'/t)^2}$.

If one were to take the bare coupling $U$ of the Hubbard model 
as the initial point of the RG scaling, the set of Eqs. 
(\ref{iu1})-(\ref{iu2}) would not provide interesting physics. 
This is so because we would have in such case $V_{intra} = 
V_{umk} = U$, and sending the cutoff 
to zero would simply reduce monotonically the 
interactions in these channels. However, the diagonals in
this space of couplings mark the boundary between the regions of
stable and unstable scaling. The slightest perturbation with
$V_{intra} < V_{umk}$ will make the
couplings to grow large, pointing at the appearance of new features 
in the system.

\begin{figure}
\epsfxsize=8cm %\epsfysize = 4cm
\centerline{\epsfbox{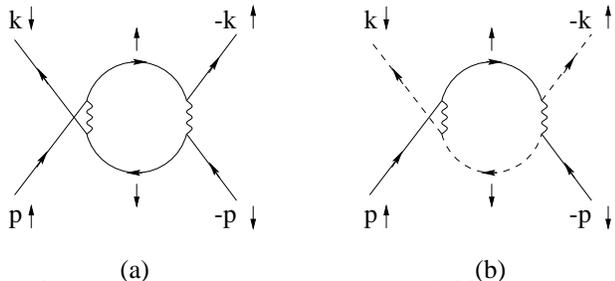}} %\vspace{-1.5cm}
\caption{Particle-hole corrections to the BCS vertices in the
Hubbard model.}
\label{three}
\end{figure}

An interesting effect comes from the fact that, even in the
Hubbard model, there are corrections that vanish in the
low-energy limit $\Lambda \rightarrow 0$ (irrelevant operators)
but may drive to the unstable regime at the early stages of
the scaling. The Kohn-Luttinger effect, that 
leads to a pairing instability in the Fermi liquid at very low
energies\cite{kl}, can be established on the same grounds in the
RG framework\cite{shankar}. 
In the case of the Hubbard model, the corrections
are given by the horizontal iteration of the bubble diagrams
in Fig. \ref{three}. The momentum flowing to the bubbles is not
in general close to $0$, in the first case, nor to
${\bf Q}$, in the second. In the wilsonian RG scheme, these
particle-hole bubbles scale therefore as $\sim (d\Lambda )^2$,
and they are considered irrelevant contributions\cite{shankar}.

In the Hubbard model with $t' < 0.276 t$, the corrections
coming from the iteration of diagram (a) in Fig. \ref{three}
have smaller strength than those from diagram (b) \cite{jpn}. 
As long as 
these are antiscreening diagrams, i.e. they add to the bare
repulsive interaction, the conditions are met to have an
unstable scaling for the vertices in Fig. \ref{two}, with the
above constraint on $t$ and $t'$. The singular behavior 
develops for the combination $V_{intra} - V_{umk}$, in which 
the particle-particle diagrams build up a pole for an
imaginary value of the frequency. The physical interpretation
is that there is a condensation of Cooper pairs at the scale 
where the vertex function diverges\cite{agd}. 
The symmetry of the order parameter
turns out to be $d$-wave, as the combination $V_{intra} - 
V_{umk}$ corresponds to having opposite amplitudes in the 
two saddle points.

\begin{figure}
\epsfxsize=\hsize %\epsfysize = 4cm
\centerline{\epsfbox{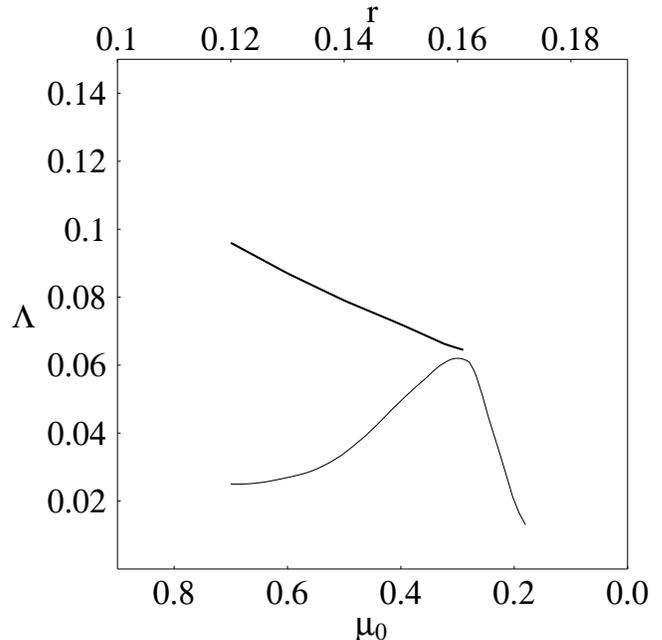}} %\vspace{-1.5cm}
\caption{Plot of the scale of the transition to the pseudogap phase
(thick line) and of the superconducting instability (thin
line) as a function of the bare chemical potential and $r = t'/t$.}
\label{four}
\end{figure}

In our approach, the scaling of the chemical potential
$\mu (\Lambda )$ regularizes the divergence of the density 
of states in Eqs. 
(\ref{iu1})-(\ref{iu2}). The result of computing the scale 
at which $V_{intra} - V_{umk}$ diverges, for a value of the
Hubbard coupling $U = 4t$, has been represented in Fig. 
\ref{four}. A singularity is only found for values of the
bare chemical potential $\mu_0 = \mu (\Lambda_0)$
which lead to attraction of the Fermi level to the VHS.
The curve of the critical scale reaches a maximum for
a certain value of optimal doping, and then it decreases for
smaller values of $\mu_0$ as the chemical potential
is not precisely pinned to the VHS in the low-energy
limit.

%We observe that, for the largest values of $\mu_0$ 
%in the pinning regime, the scale of the superconducting
%transition is relatively small. This is due to the fact that 
%the irrelevant contributions that trigger the superconducting
%instability have small strength at the early stages of the RG
%process, when the chemical potential is not close to the VHS.
%The curve of the critical frequency reaches a maximum for 
%a certain value of optimal doping, and then it decreases for
%smaller values of $\mu_0$ as the chemical potential 
%is not precisely pinned to the singularity in the low-energy 
%limit. 

The instability in the BCS channel has to be matched against
the strong tendency towards a magnetic instability at wave-vector 
${\bf Q} = (\pi, \pi )$ for $t' < 0.276 t$ \cite{jpn}. 
In the RG framework, this comes from
the existence of particle-hole contributions which grow large 
at low energies. They are built from diagrams 
similar to (b) of Fig. \ref{three}, but with the momentum  
flowing into the bubble being precisely ${\bf Q}$. This kinematics
corresponds to the vertices $E_{\bf Q}$ and 
$E_U$ that we have defined in Fig. \ref{twop}.

It can be seen that, at least at the one-loop level, the 
vertices $E_{\bf Q}$ and $E_U$
are only renormalized by interactions with their own kinematics
in the wilsonian approach. This leads to the scaling equation
\begin{equation}
\Lambda \frac{\partial (E_{\bf Q} + E_U)}{\partial \Lambda}
    =   - c' (E_{\bf Q} + E_U)^2  /(4\pi^2 t)
\label{exch}
\end{equation}
with $c' \equiv \log \left[ \left(1 + \sqrt{1 - 4(t'/t)^2} 
\right)/(2t'/t) \right]$.
In the low-energy limit, a singularity is reached at a 
certain value of $\Lambda $. The logarithmically
divergent particle-hole
bubbles building up the singularity have an imaginary 
part equal to $i\pi /2$ times $c'/(4\pi^2 t)$, which means
that the vertex gets actually a pole for an imaginary 
value of the frequency.

\begin{figure}
\epsfxsize=8.5cm %\epsfysize = 4cm
\centerline{\epsfbox{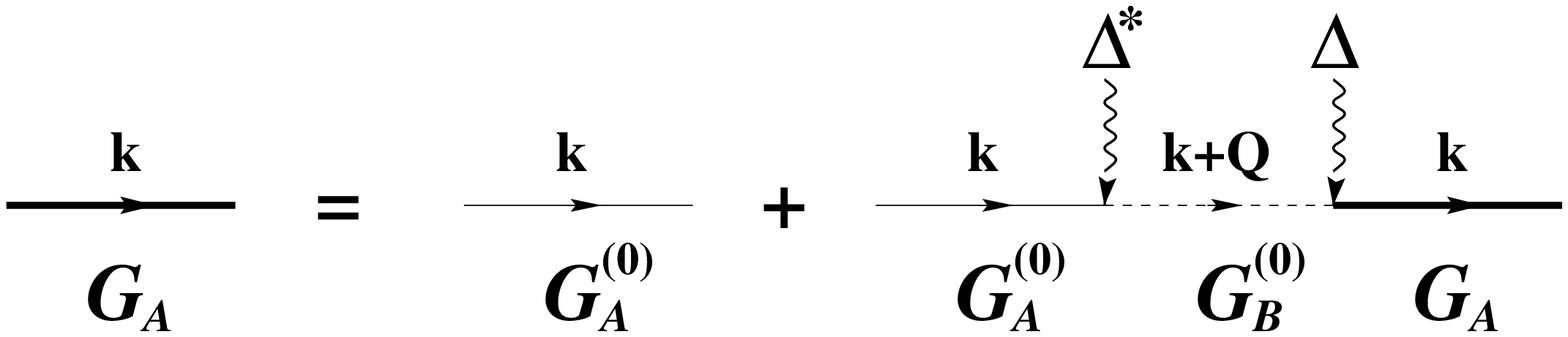}} %\vspace{-1.5cm}
\caption{Self-consistent equation for the dressed electron   
propagator $G$ in the particle-hole condensate, in terms of 
the undressed propagators $G_A^{(0)}$
and $G_B^{(0)}$ at the two inequivalent saddle points.}     
\label{five}
\end{figure}

Thus, the divergence in this channel gives rise to 
the condensation of particle-hole pairs with momentum 
${\bf Q}$. This physical interpretation is similar to that 
of the superconducting instability\cite{agd}, 
but the presence in this case of a macroscopic
number of particle-hole pairs leads to nonvanishing
expectation values of the type
$\int d^3 k \langle \Psi^{+}_{A\uparrow} ({\bf k})
\Psi_{B\downarrow} ({\bf k}+{\bf Q}) \rangle \equiv \Delta $,
$\Psi_{\alpha \sigma} ({\bf k})$ being the electron field 
operator at saddle point $\alpha $. This has a drastic 
effect in the low-energy spectrum, 
since the electron propagator is corrected by the insertion
of the condensate as shown in Fig. \ref{five}. The result
is that a gap of magnitude $|\Delta |$ opens up in the spectrum
of quasiparticles. The appearance of this gap takes place
in the neighborhood of the saddle points, while gapless
quasiparticle excitations still exist in the rest of the
Fermi line not affected by the nesting of such singular 
points.

The point at which the vertex $E_{\bf Q} + E_U$ diverges can be 
estimated from Eq. (\ref{exch}), and it has been represented 
as a function of $t'$ in Fig. \ref{four}.
In accordance with the known results about the renormalization 
of $t'$ with doping\cite{ogata}, we have assumed a variation of 
its effective value following the qualitative trend shown 
in the figure.
The scale of condensation of particle-hole pairs is larger
than the scale of the superconducting instability up to the
point of optimal doping marked by the peak of the latter.
In this regime, the opening of a gap of magnitude $|\Delta |$
is the effect that prevails, and the renormalization of the
interactions is stopped at that energy scale.

On the other hand, the superconducting instability takes over
beyond the optimal doping. The chemical potential departs 
then from the VHS at low energies, by an amount that becomes 
of the same order as the scale that would be predicted for
the magnetic instability. In these conditions the scaling of 
$E_{\bf Q} + E_U$, which relies on the
pinning to the VHS, is arrested before the singularity in
the vertex is reached. This is in contrast to the scaling in 
the BCS channel given by Eqs. (\ref{iu1})-(\ref{iu2}), which
holds irrespective of the pinning to the VHS \cite{shankar}.

A closer look at the origin of the singularity in the
particle-hole sector shows that it arises without a
preferred direction of the spin. The divergence in
the exchange channel may result in a nonvanishing expectation 
value with spin projection $S_x$ or $S_y$, depending on the 
choice of the condensate wavefunction. A similar divergence 
in the channel with momentum transfer ${\bf Q}$ and no 
spin-flip shows that
$\int d^3 k \langle \Psi^{+}_{A\uparrow} ({\bf k})     
\Psi_{B\uparrow} ({\bf k}+{\bf Q}) \rangle $ has also an
absolute value precisely equal to $|\Delta |$. Depending on 
the different combinations of phases, the spin of the 
condensate may point in any direction of the space, what 
provides a check of the rotational invariance of the wilsonian 
RG scheme.

When considering the model strictly at zero temperature,
the ground state of the system is forced to choose a definite 
projection of the spin. The condensation of particle-hole pairs 
leads then to the spontaneous breakdown of the $SO(3)$ 
invariance. As a consequence, a pair of Goldstone bosons arise 
corresponding to spin waves on top of the condensate of
particle-hole pairs. In the regime below optimal doping where 
the magnetic instability prevails, these are the gapless
excitations of the spectrum together with the quasiparticles 
from the regions
not affected by the nesting of the saddle points.

We have thus clarified the nature of the magnetic instability
that arises when the Fermi level is pinned to the VHS of 
the electron system. The main physical effect is the 
condensation of particle-hole pairs, which results in the 
opening of a gap for the quasiparticles in the neighborhood
of the saddle points. 
This is the dominant instability up to the optimal doping
marked by the peak of the scale of the superconducting 
instability, which takes over for higher doping levels.
The properties that we have discussed
rely on the existence of an attractive fixed-point in the
scaling of the chemical potential of the 2D system. They
are therefore robust as they do not depend on fine-tuning
or on the particular details of the model, what may
explain some of the universal properties of the 
hole-doped copper-oxide superconductors.

%We have seen that the most prominent 
%experimental features of the cuprates may be explained by
%the proximity of the Fermi level to a VHS in the 2D 
%layers. The picture that emerges from Fig. \ref{four} is 
%consistent with the phase diagram in the doped regime 
%of the copper-oxide superconductors.
%As pointed out elsewhere\cite{np}, features like
%the attenuation of the quasiparticle weight are also 
%inherent to the behavior of electrons at a saddle-point
%dispersion relation and in agreement with phenomenological
%models of the cuprates\cite{varma}. 
%The properties that we have discussed
%rely on the existence of an attractive fixed-point in the
%scaling of the chemical potential of the 2D system. They
%are therefore robust as they do not depend on fine-tuning 
%or on the particular details of the model, what may
%explain in turn the universal properties of the wide class
%of hole-doped copper-oxide superconductors.   

Fruitful discussions with F. Guinea 
are gratefully acknowledged. This work has been partly 
supported by CICyT (Spain) and CAM (Madrid, Spain) 
through grants PB96/0875 and 07N/0045/98.

\end{document}